\documentclass[useAMS,usenatbib]{mn2e}
\usepackage{graphicx}
\usepackage{color}
\usepackage{float,lscape}
\usepackage{amsmath}
\usepackage{amssymb}

\title[Gaia SPSS variability monitoring]{The Gaia spectrophotometric standard
stars survey --- III. Short-term variability monitoring\thanks{Based on data
obtained within the Gaia DPAC (Data Processing and Analysis Consortium) --- and
coordinated by the GBOG (Ground-based Observations for Gaia) working group ---
at various telescopes; see acknowlegements.}}
\author[S. Marinoni et al.]
{S.~Marinoni$^{1,2}$\thanks{E-mail:silvia.marinoni@asdc.asi.it}, 
 E.~Pancino$^{3,2}$,
 G.~Altavilla$^{4}$,
 M.~Bellazzini$^{4}$,
 S.~Galleti$^{4}$, 
 \newauthor
 G.~Tessicini$^{4}$,
 G.~Valentini$^{5}$,
 G.~Cocozza$^{4}$,
 S.~Ragaini$^{4}$,
 V.~Braga$^{6,2}$
 \newauthor
 A.~Bragaglia$^{4}$,
 L.~Federici$^{4}$,
 W.J.~Schuster$^{7}$,
 J.~M.~Carrasco$^{8}$,
 A.~Castro$^{7}$,
 \newauthor
 F.~Figueras$^{8}$,
 and C. Jordi$^{8}$\\
$^{1}$INAF-Osservatorio Astronomico di Roma, Via Frascati 33, I-00040, Monte Porzio Catone (Roma), Italy\\
$^{2}$ASI Science Data Center, via del Politecnico SNC, I-00133 Roma, Italy\\
$^{3}$INAF-Osservatorio Astrofisico di Arcetri, largo Enrico Fermi 5, I-50125 Firenze, Italy\\
$^{4}$INAF-Osservatorio Astronomico di Bologna, via Ranzani 1, I-40127 Bologna\\
$^{5}$INAF-Osservatorio Astronomico di Teramo, via Mentore Maggini SNC, I-64100 Teramo, Italy\\
$^{6}$Department of Physics, Universit\`a di Roma Tor Vergata, via della Ricerca Scientifica 1, 00133 Roma, Italy\\
$^{7}$Observatorio Astron\'omico Nacional, Universidad Nacional Aut\'onoma de M\'exico, Apartado Postal 877,
            C.~P.~22800 Ensenada, B.~C., M\'exico\\
$^{8}$Departament de F\'isica Qu\`antica i Astrof\'isica, Institut del Ci\`ences del Cosmos (ICC), Universitat de Barcelona (IEEC-UB), \\
            c/ Mart\'\i\  i Franqu\`es, 1, 08028 Barcelona, Spain\\
}
\begin{document}

\date{Accepted ... Received .... ; in original form ....}

\pagerange{\pageref{firstpage}--\pageref{lastpage}} \pubyear{2016}

\maketitle

\label{firstpage}

\begin{abstract}

We present the results of the short-term constancy monitoring of candidate Gaia
Spectrophotometric Standard Stars (SPSS). We obtained time series of typically
1.24~hour -- with sampling periods from 1--3~min to a few hours, depending on the
case -- to monitor the constancy of our candidate SPSS down to 10~mmag, as
required for the calibration of Gaia photometric data. We monitored 162 out of a
total of 212 SPSS candidates. The observing campaign started in 2006 and
finished in 2015, using 143 observing nights on nine different instruments covering both hemispheres.
Using differential photometry techniques, we built light curves with a typical precision of 4~mmag, depending
on the data quality. As a result of our constancy assessment, 150 SPSS candidates 
were validated against short term variability, and only 12 were rejected because of variability 
including some widely used flux standards such as BD+174708, SA 105-448, 1740346, and HD 37725.

\end{abstract}



\begin{table*}
\centering
\caption{Observation diary of the constancy monitoring campaign. The columns
contain: {\em (1)} the instrument used; {\em (2)} the telescope; {\em (3)} the
observing site; {\em (4)} the campaign time span; {\em (5)} the number of nights awarded to the short-term
monitoring campaigns (within brackets the number of useful nights); {\em (6)} the
number of observed time series; {\em (7)} the number of monitored
SPSS. \label{tab:obsdiary}}
\begin{tabular}{@{}llllcccl@{}}
\hline
Instrument &  Telescope &   Site & Period  & \# of (useful) nights & \# of Light Curves & \# of SPSS  \\
\hline
BFOSC      &  Cassini   & Loiano              & Aug 2006 -- Jul 2015   & 33 (20)  & 119   &  85 \\
DOLORES    &  TNG       & La Palma            &	May 2007 -- Aug 2014   &  9  (5)  &  29   &  27 \\
ALFOSC     &  NOT       & La Palma            &	Apr 2014 -- Mar 2015   &  1  (1)  &   4   &   4 \\
CAFOS      &  2.2m      & Calar Alto          & Sep 2008 -- May 2012   &  7  (4)  &  21   &  17 \\
EFOSC2     &  NTT       & La Silla            &	Apr 2009 -- Jan 2015   &  3  (3)  &  19   &  17 \\
LaRuca     &  1.5m      & San Pedro M\'artir  & Jan 2008 -- May 2011   & 33 (19)  & 113   &  86 \\
ROSS       &  REM       & La Silla            &	Jul 2007 -- Jul 2012   & 48 (27)  & 124   &  76 \\
ROSS2      &  REM       & La Silla            &	Feb 2015 -- Jun 2015   &  9  (7)  &  31   &  30 \\
MEIA       &  TJO       & Montsec             & Jun 2009 -- May 2013   & 58 (24)  &  32   &   2 \\
\hline
\end{tabular}
\end{table*}

\section{Introduction\label{sec:intro}}

The Gaia\footnote{http://www.cosmos.esa.int/web/gaia/home} satellite \citep{gaia1,gaia2} is a
cornerstone mission of the ESA (European Space Agency) Space Program, launched
by a Soyuz-Fregat vehicle in 2013 December 19 from the European
Spaceport in Kourou, French Guiana.  Gaia is performing an all-sky survey to obtain positions,
parallaxes and proper motions to $\mu$as precision for more than one billion
point-like sources on the sky. Expected accuracies are in the 7--25 {\it
$\mu$as} range down to 15-th mag and sub-mas accuracies down to a limiting
magnitude of V $\simeq$ 20 mag.  The astrometric data are complemented by 
low-resolution spectrophotometric data in the 330--1050 nm wavelength range and,
for the brightest stars ($V < 16$ mag), by radial velocity measurements  in the
spectral region centered around the calcium triplet (845--872 nm) at a resolution of about
$R=\lambda/\Delta\lambda \simeq$11500, with 1--15 km~s$^{-1}$ errors, depending on
spectral type and brightness.

The Astrometric Field CCDs will provide G-band images, i.e., white light images
where the passband is defined by the telescope optics transmission and the CCD
sensitivity,  with a very broad combined passband ranging from 330 to 1050 nm
and peaking around 500--600 nm.  The blue (BP) and red (RP) spectro-photometers
will provide dispersed images with 20$<\lambda/\Delta\lambda<$100 over the
spectral ranges 330--680 nm and 640--1050 nm, respectively.

The final conversion of internally-calibrated G instrumental magnitudes and
BP/RP instrumental fluxes into physical units requires an external absolute
flux scale, that our team is in charge of providing \citep{pancino12}. Ideally, the 
Gaia spectrophotometric standard stars (SPSS) grid should comprise of the
order of 200 SPSS in the range $9\le V\le15$~mag, properly distributed in the
sky to be observed by Gaia as many times as possible. The mission requirement
is to calibrate Gaia data with an accuracy of a few percent (1--3\%) with
respect to Vega \citep{bohlin07}.

The obvious and fundamental requirement for a SPSS is  that its magnitude
(flux) is constant. Since only a few of our SPSS candidates have accurate enough photometry
in the literature and have been monitored for variability in the past, it is  very important to perform
repeated and accurate observations of our  targets in order to detect
variability larger than 0.01 mag (if any), both intrinsic (e.g., pulsations) or
extrinsic (i.e., binarity),  before we invest a large amount of resources and
observing time in getting their absolute SEDs. In addition, we remind that even
stars used for years as spectrophotometric standards have been found to vary
when dedicated studies have been performed \citep[see, e.g., G24-9, that has
been found to be an eclipsing binary,][]{landolt07}.

Additionally, a comprehensive search of the literature for each candidate analyzed 
in this research was conducted. Most of our SPSS are white-dwarfs (WD hereafter) and  hot subdwarfs, the
remaining are dwarf/giant stars covering different spectral types, including
cool stars  of late spectral type up to M.  WDs may show variability with
(multiple) periods from about 1 to 20 min and amplitudes from  about 1--2\% up
to 30\%. We have tried to exclude stars within the instability strips for white dwarfs 
of type ZZ Ceti, DQV, V777 Her or GW Vir \citep[see][and references therein]{castanheira07,althaus10}. 
However, in many cases the existing
information  is not sufficient (or sufficiently accurate) to firmly establish
the constant nature of a given WD. Hence, many of our WD SPSS
candidates needed to be monitored. Similar considerations are valid for hot
subdwarfs \citep{kilkenny07}. 

Redder stars are also often variable: K stars have shown variability of 5--10\%
with periods of the order of days to tens of days \citep{eyergrenon97};  M
stars can vary, for example, because of flares. In addition, binary systems are
frequent and eclipsing binaries can be  found at all spectral types. The
periods of the known variables span from a few hours to hundreds of days, most of them
having $P\sim$ 1--10 days \citep{dvorak04}.

The main use of the SPSS grid is of course the absolute calibration of Gaia
spectrophotometric data. Nevertheless, such a large grid (more than 200 stars)
represents  an unprecedented catalogue of spectro-photometric standard stars,
characterized by high precision and accuracy, full sky coverage, including stars
spanning a wide  range of spectral types, and with spectra covering a large
wavelength range. For comparison,
CALSPEC\footnote{http://www.stsci.edu/hst/observatory/crds/calspec.html}
\citep{bohlin07}, containing the composite stellar spectra which are the
fundamental flux standards for HST calibrations, consists of about 90 stars: our
SPSS  grid will be more than two times larger, and it  will be about two times
larger than the \citet{stritzinger05} catalogue. Our sample will be comparable
to  recent catalogues, such as the STIS Next Generation Spectral Library Version
2\footnote{http://archive.stsci.edu/prepds/stisngsl/} which, in any case, does not
meet as well all Gaia requirements (such as, for example, the magnitude
range).\\

This paper is the third of a series \citep[see][hereafter Paper~I and Paper~II,
respectively]{pancino12,altavilla15}, and presents the results of our short-term
constancy monitoring campaign.  It is organized as follows: the observation
strategy and data reduction are briefly described in Section~\ref{sec:obsred};
Section~\ref{sec:relphot} describes the light curves production and our quality
control (QC) procedures, and  describes our criteria for constancy assessment;
the results of the short-term constancy monitoring
campaigns are presented and discussed in Section~\ref{sec:res}. We summarize our
results and present our conclusions in Section~\ref{sec:conclusion}.

\section{Observations and data reduction\label{obsred}}
\label{sec:obsred}

The various observing campaigns for the absolute flux calibration of the Gaia
spectrophotometry (Paper~I), are both spectroscopic and photometric. At survey
completion, the SPSS observing campaings amounted to 515
observing nights, at eight different telescopes and nine instruments (see Table
\ref{tab:obsdiary}), with 58\% useful data, and the rest lost for bad weather,
technical problems, or other reasons. The global campaign produced approximately
102\,500 frames to be quality controlled, reduced, and analyzed. The survey
started in August 2006 and ended in July 2015.

\subsection{Photometry observations}

Photometry observations were performed in two different flavours: {\em night
points} or {\em time series}. The night point can be absolute or relative
(depending on the sky conditions), and is formed by at least 9 consecutive
frames acquired in the Johnson-Cousins B, V and R filters  (and sometimes also
I and U). The study of SPSS absolute night points will be presented in a future
dedicated paper (see Paper~I for more details about the observing strategy details).

\begin{figure}
\centering
\includegraphics[angle=270,width=\columnwidth]{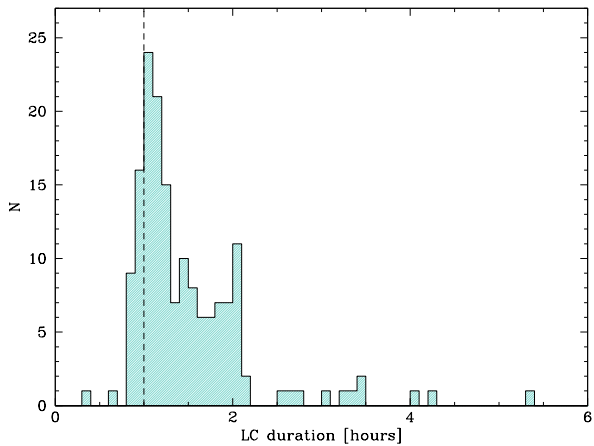} 
\caption{Distribution of time series durations for the 162
SPSS presented in this paper. A few series last less than 1~h (vertical dashed
line), but the typical duration is 1.26~h with some curves lasting as long as
5.4~h.} 
\label{fig:duration}
\end{figure}

A time series contains at least 30 consecutive exposures in one filter (normally
in the bluest available) covering approximately  1--2 hours, in order to
monitor the short-term photometric constancy for each SPSS candidate. The
actual duration of the 162 time series presented in this paper is illustrated in
Figure~\ref{fig:duration}. We will focus here on the analysis of time
series, in order to validate our SPSS candidates against short-term variability
phenomena.  Data coming from the night points obtained close to a time series
were sometimes included, when useful and appropriate. We also use our preliminary results 
on the absolute night points to further investigate the long-term constancy of our SPSS candidates 
(see Section.~\ref{sec:res}).

The candidate SPSS were prioritized according to available literature data,
placing them more or less close to the main instability strips and variability
regions across the parameter space. They were then observed, subject to the
scheduling allocation of various observatories, and to visibility and weather
constraints. The short-term constancy monitoring campaign shared the allocated
time with the absolute photometry and the spectroscopy campaigns --
approximately 143 nights were dedicated to it -- and as a result was not always
performed in optimal conditions. 
Consequently, not all candidate SPSS were
observed, and some were observed with non-optimal data\footnote{Of the 212 initial 
candidates reported in Paper~I, for 43 the kind of monitoring described 
in the present paper was not considered necessary because, according to the 
available literature data the SPSS candidate was not close to any of the instability regions. 
Of the remaining list of 169 prioritized candidates for short 
term monitoring, 162 could be observed with the described strategy and for 
7 we did not obtain usable data.}. In addition to our study, 
Gaia observations themselves will further check variability, because each SPSS 
will be observed tens of times. Hidden, nearby companions may be discovered 
from space. These stars will simply be eliminated from the SPSS grid, which is built with some redundancy.

As can be seen from
Table~\ref{tab:obsdiary}, each SPSS candidate was observed multiple times (on average,
roughly 3 series per SPSS), either to repeat a non-optimal series, or to
observe a subset of SPSS with different telescopes to compare the results. Here
we present only the best time series obtained for each star.

\subsection{Data reduction}

\begin{table*}
\centering
\caption{Best light curves for the 162 monitored SPSS. The columns contain: {\em (1)} the internal SPSS 
ID; {\em (2)} the SPSS name; {\em (3)} the instrument used; {\em (4)} the Heliocentric Julian day; 
{\em (5)} the photometric band; {\em (6)} the magnitude difference of each point with respect
to the average magnitude of comparison stars; and {\em (6)} the
errors of individual epoch measures for the SPSS (see text for more details). The relative photometry 
table is published in its entirety in the electronic online version of the
Journal, and at CDS. Here we show just a few lines to illustrate its contents.
\label{tab:curves}}
\begin{tabular}{@{}lllcccc@{}}
\hline
SPSS   & Star name    & Instrument &  HJD  & Band & $\Delta$m & $\delta_m$   \\
       &              &            & (day) &      &   (mag)   &   (mag)      \\
\hline
042  &  P~41-C          &  BFOSC   &  2454877.67853892  & B &  -0.0000 & 0.0013 \\
042  &  P~41-C          &  BFOSC   &  2454877.68073774  & B &   0.0010 & 0.0015 \\
042  &  P~41-C          &  BFOSC   &  2454877.68225397  & B &   0.0025 & 0.0014 \\
042  &  P~41-C          &  BFOSC   &  2454877.68536727  & B &  -0.0013 & 0.0016 \\
042  &  P~41-C          &  BFOSC   &  2454877.68676764  & B &   0.0017 & 0.0015 \\
\hline
\end{tabular}
\end{table*}

To ensure that the maximum quality could be obtained from the SPSS photometric
observations, a careful data reduction protocol \citep{marinoni12}  was
implemented, following an initial assessment \citep{marinoni11,marinoni13} of
all the instrumental effects that can have an impact on the photometry
precision and accuracy. The methods and results of such instrumental effects
study were presented in their final form in Paper~II. Particularly relevant for
the present paper are the characterizations of the {\em minimum acceptable exposure
time}, the CCD linearity study, and the automated quality assessment and
stability monitoring of the calibration frames (on timescales as long as 9
years in some cases). 

The data reduction methods were fairly standard but, due to the large amount of
data collected, an automated IRAF-based\footnote{IRAF is the Image Reduction
and Analysis Facility, a general purpose software system for the reduction and
analysis of astronomical data. IRAF is written and supported by the IRAF
programming group at the National Optical Astronomy Observatories (NOAO) in
Tucson, Arizona. NOAO is operated by the Association of Universities for
Research in Astronomy (AURA), Inc. under cooperative agreement with the
National Science Foundation.} pipeline was built in order to take care of both
the quality control of the images and the removal of all the instrumental
signatures. We applied the usual detrending steps including dark removal (for
REM and MEIA), bias and overscan (when available) correction, flat fielding, bad
pixel correction, and fringing correction when relevant. 

\subsection{Data availability}

The raw and reduced data are stored at ASDC (ASI Science Data
Center\footnote{http://www.asdc.asi.it/}) in a dedicated database that will
contain all the data products of the SPSS campaign and will be opened to the
public with the first SPSS public data release in the near future.
Additionally, the light curves presented in this paper will be published in the
electronic version of the Journal and in the CDS Vizier
service\footnote{http://vizier.u-strasbg.fr/}, with the form illustrated in
Table~\ref{tab:curves}.

\begin{figure}
\centering
\includegraphics[angle=270,width=\columnwidth]{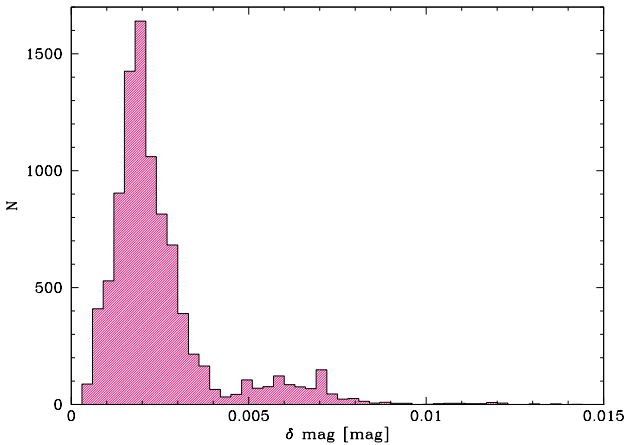} 
\caption{Histogram of the errors on the differential SPSS magnitudes for each epoch,
based on SExtractor formal errors on the aperture magnitudes. The histogram
contains 9377 single-epoch catalogues, used to build the best light curves presented in
this paper.} 
\label{fig:error}
\end{figure}


\section{Light curves}
\label{sec:relphot}

We describe here the procedure to obtain relative light curves from the
aperture magnitudes measurements on the single frames, the QC process, and the
 adopted criteria for constancy assessment.

\subsection{Aperture photometry}
\label{sec:catsandQC}

Magnitudes were measured with SExtractor \citep{Bertin1996}, a simple and powerful tool to
perform reliable photometric measurements. We used the variety of
flags and parameters output by the code, to write the
semi-automated procedures for the QC of our large data set. 

Our observed fields are not crowded, and we aimed at very high
precision (10~mmag or better), therefore we performed aperture photometry. 
We were free to choose an aperture large enough to avoid significant light losses,
therefore we generally used six times the FWHM (Full Width at Half Maximum)
that granted light losses well below 1\%. 

The SExtractor catalogues were cross-matched with the catalogue
cross-correlation software CataPack\footnote{The CataPack package is developed
by P.~Montegriffo at the Bologna Observatory (INAF).  CataXcorr and
Ca\-ta\-Comb are parts of this package, which is available at
http://www.bo.astro.it/$\sim$paolo/Main/CataPack.html.}, and in particular with
the CataXcorr routine, to cross-identify the SPSS candidate and the comparison stars in each
frame of a time series. A final catalogue was then created with the CataComb
routine, containing the aperture magnitudes of the target SPSS and of suitable
comparison stars in the field, that should be visible in all the frames included
in a series.
We selected at least two bright stars around the SPSS, and
we monitored the magnitude difference between the target and the comparison stars, 
as described more in detail in Section \ref{sec:LCanal}.
If one of the comparison stars was found to be variable, it was rejected from the 
adopted set of comparison stars.

The errors on the differential SPSS magnitudes ($\delta$mag) -- based on the
formal SExtractor errors on the aperture magnitudes for each epoch -- are displayed in
Figure~\ref{fig:error}. The histogram contains 9377 single-epoch catalogues, used to
build the best light curves presented in this paper. As can be seen, the typical
error is 0.002~mag, with some measurements reaching as high as 0.01~mag and more, but also many 
going as low as 0.0005~mag.

Instrumental effects were investigated in detail in Paper~II, as mentioned previously, 
and taken into account as explained there. Briefly, we decided to exclude from the present 
analysis all the images affected by low S/N, saturation, non-uniform CCD illumination 
or significant geometric distortions. A few dubious cases in which atmospheric conditions 
were not optimal are discussed in Section~\ref{sec:res}. Moreover, the use of strictly differential photometry
ensures that all remaining systematic or instrumental error soruces are beaten below the 1\% level -- which is
our requirement for Gaia -- and therefore were not explicitly considered in the error computation.

\subsection{Quality control}
\label{sec:qcprocedure}

Because we required a high precision in the final light curves, great care was
taken in the selection of appropriate frames and time series for the analysis,
with three levels of automated QC. At the {\em star level}, both the SPSS and
the candidate comparison stars measurements had to satisfy a set of criteria: 
SNR$>$100, no saturation (using the measurements presented in
Paper~II), non-distorted PSF, seeing below 5", and no bad pixels within the
selected aperture. At the {\em frame level}, we rejected frames in which the
SPSS did not pass the star level QC, or where there were less than two
comparison stars passing the star level QC. Finally, at the {\em series level},
we only accepted time series with at least 30 exposures passing the frame level
QC, and lasting in total at least one hour. 

The QC procedure issues warnings that can be examined later to assess the
causes. In fact, our QC criteria are quite restrictive, and in some cases we
were forced to accept some series that did not actually pass all of the QC
chain. For example, when no better series existed for a particular SPSS, we
accepted series slightly shorter than 30 points, or with one comparison star
only, or with a large spread in the average comparison star magnitudes. We used these less
precise data anyway, to put some constraints on the maximum variation
amplitude allowed by the data, on the tested timespan of 1-3~hours, as further discussed in 
Sections \ref{sec:constass} and \ref{sec:res}.

The actual duration of the 162 time series presented in this
paper is illustrated in Figure~\ref{fig:duration}. It can be seen that a few
light curves shorter than 1~hour were included for lack of better data, while
the typical duration is 1.26~h, with a few curves lasting 4~h or more.

\begin{figure}
\centering
\includegraphics[width=\columnwidth]{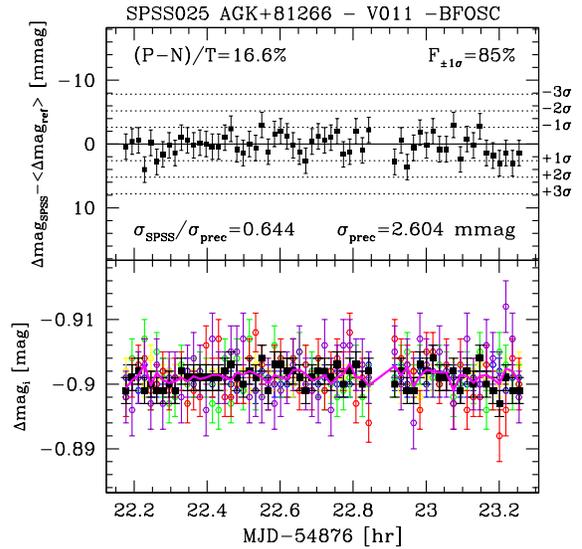} 
\caption{An example of checkplot produced by our light curve pipeline. The star
is AGK+81266, our SPSS~025, observed from Loiano with BFOSC@Cassini, in our run
V-011 (February 2009). In the {\em bottom panel} are the normalized light
curves of the SPSS (in black) and of a few comparison stars (each in a
different colour), as a function of the Modified Julian Day (MJD). The
super-comparison light curve is also plotted as a magenta curve. In the {\em
top panel} the zeroed light curve for the SPSS is plotted, along with some
useful quantities for light curve validation and SPSS constancy assessment (see
text for more details). The dotted lines mark the $\pm$1, 2, and
3~$\sigma_{\rm prec}$ thresholds.} 
\label{fig:LCcompleteexample}
\end{figure}

\subsection{Light curves production} 
\label{sec:LCanal}

Using the catalogues that passed the QC criteria already described in Section~\ref{sec:qcprocedure}, we built the
light curves in two main steps. In the first step we created a {\em
super-comparison} light curve using all the chosen comparison stars. At each
epoch, the average flux of the chosen comparison stars was computed and converted
into an average epoch magnitude. For each star, the difference
between its epoch magnitude and the average epoch magnitude above was then used to build a light
curve. The comparison light curves were reported to the same zeropoint using
the difference between the median magnitude of each curve with that of a chosen
comparison curve, usually the one of the star with the highest S/N ratio (see bottom panel 
of Figure~\ref{fig:LCcompleteexample}).
The super-comparison curve (magenta line in Figure~\ref{fig:LCcompleteexample}) was then created as the average 
of normalized light curves for all the chosen comparison stars. 

As a second step, the SPSS light
curve was computed, normalized to zero. This was achieved by first computing
the SPSS normalized light curve in the same way as for the comparison stars (black filled circles in the bottom panel of 
Figure~\ref{fig:LCcompleteexample}) and then by subtracting the super-comparison curve from it (top panel of 
Figure~\ref{fig:LCcompleteexample}).

A number of useful quantities were also computed, along with a checkplot that
summarizes all relevant information (Figure~\ref{fig:LCcompleteexample}). The quantities that were used for
the light curve QC and constancy assessment are the following:

\begin{itemize}
\item{$\sigma_{\rm prec}$ is the average standard deviation of the reference stars and tracks
the precision to which the magnitude of constant stars is reproduced over the
whole series;}
\item{$\sigma_{\rm SPSS}$ is the standard deviation of the SPSS light curve;}
\item{the ratio $\sigma_{\rm SPSS}/\sigma_{\rm prec}$ is an indication of variability:
when this ratio is above one, the variation detected in the SPSS light curve is
larger than the curve precision; there is a significant indication of 
variability when the ratio becomes higher than 3, for example;} 
\item{$F_{\pm1\sigma}$ is the fraction of SPSS light curve points
lying within $\pm$1$\sigma_{\rm prec}$; similarly to the ratio described above,
this is another way of searching for variability: when the fraction is significantly below
68\%, the distribution of points is not normal;}
\item{finally, $(P-N)/T$ is a fractional indicator of asymmetry in the
distribution of points, where $P$ is the number of points above zero, $N$ below
zero, and $T$ the total number of points in the curve.}
\end{itemize} 

\begin{figure}
\centering
\includegraphics[angle=270,width=\columnwidth]{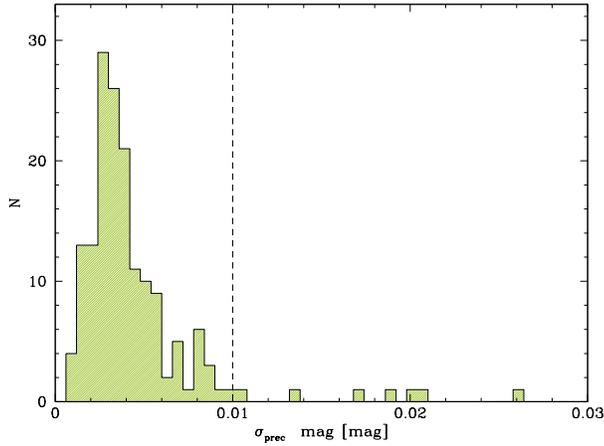} 
\caption{Histogram of $\sigma_{\rm prec}$ of the 162 light
curves presented in this paper. The vertical dashed line marks our criterion of
0.01~mag. As can be seen, the typical curve precision is of 0.004~mag and there
are only 7 light curves that do not match our criterion, and that
were included for lack of better data.}
\label{fig:sigma}
\end{figure}

After light curve production, a simple QC criterion is applied: $\sigma_{\rm prec}
\leq$10~mmag. Ideally, constancy assessment is relevant for Gaia only when
based on light curves passing this criterion. In practice, as mentioned
previously, for some SPSS we also accepted curves that did not pass the
criterion, for lack of better data, because we assumed that even some less
strict indication of the SPSS nature was better than no indication at all. 
The distribution of $\sigma_{\rm prec}$ for the 162 light curves
presented in this paper is shown in Figure~\ref{fig:sigma}, where it can be
noticed that 7 light curves failing this QC step were
included. We note that the typical precision of our light curves is comparable to that 
of recent studies for the search of exoplanets from space 
\citep{nascimbeni12} or from the ground \citep{klagyivik16}.

An output file containing the resulting (relative) light curve for
each time series of each SPSS was produced and archived at ASDC. The best light
curve for each SPSS, presented in this paper, is also published electronically 
in the form presented in Table~\ref{tab:curves}.

\subsection{Constancy assessment}
\label{sec:constass}

The simplest form of variable star analysis is the inspection of the shape of
the light curve, and the time and magnitude of maximum and minimum, if any. The
term {\it range} is here adopted to denote the difference between maximum and
minimum (if present), or the maximum magnitude excursion observed in the curve if
no clear maximum and minimum are detected. The term {\it amplitude} ($A$) is used to denote 
the half range, as in the coefficient  of a sine or cosine function. 

\begin{figure}
\centering
\includegraphics[angle=270,width=\columnwidth]{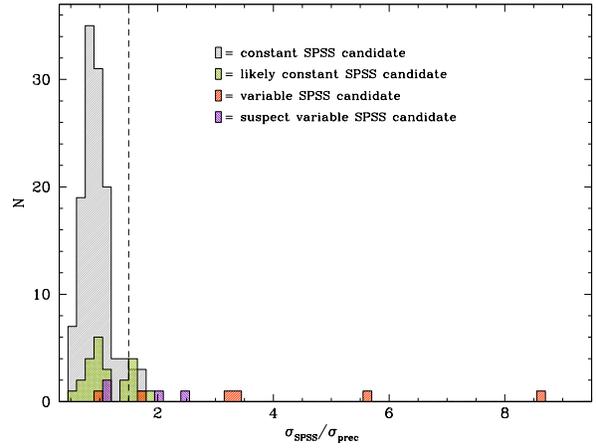} 
\caption{Histogram of $\sigma_{\rm SPSS}/\sigma_{\rm prec}$ of the 162 light curves
presented in this paper. The vertical dashed line marks our approximate limit
for constancy. The constant stars are highlighted in grey, the likely constant
ones in green, the suspect variable stars in purple, and the confirmed variable
stars in orange.} 
\label{fig:ivar}
\end{figure}

For the light curves passing our QC criterion ($\sigma_{\rm prec} \leq 10$~mmag),
we contemplate two cases:

\begin{enumerate}
\item{a candidate SPSS is validated as {\em constant} over the sampled periods
if no coherent pattern is present in the light curve and if
$\sigma_{\rm SPSS}/\sigma_{\rm prec} \lesssim 1.5$\footnote{This is not
a strict limit, but all stars close to this value were checked more carefully
than the ones far away from it. A posteriori, we can say that the median value
of $\sigma_{\rm SPSS}/\sigma_{\rm prec}$ is 0.91 for the constant stars, 1.05 for the
likely constant stars, 1.97 for the suspect variables, and 5.64 for the
confirmed variables, as illustrated in Figure~\ref{fig:ivar}.}; in that case,
even if the star varies, its amplitude is A$<\sigma_{\rm SPSS}\pm\sigma_{\rm prec}$ and
it is accepted as an SPSS;} 
\item{a candidate SPSS is judged {\em variable} and rejected as an SPSS if a
coherent pattern is evident; in this case, we define $A_{max} =
max(LC_{max},|LC_{min}|)$, where $LC_{max}$ and $LC_{min}$ are the maximim and minimum magnidude differences measured on the LC,
and if A$_{max}>3\sigma_{\rm prec}$ the pattern is
significant; the estimated amplitude of the variation is:}
\begin{itemize}
\item{if no clear maximum or minimum is visible in the curve, the period of the
variation is most probably longer than the time series, and we assumed that
A$>$A$_{max}\pm\sigma_{\rm prec}$;}
\item{if only the maximum or minimum is visible in the curve, the
period of the variation is again most probably longer than the time series, but
we assumed that A$\geq$A$_{max}\pm\sigma_{\rm prec}$;}
\item{if both the maximum and minimum are visible, we assumed
A=A$_{max}\pm\sigma_{\rm prec}$, but of course this is only valid over the sampled
periods.}
\end{itemize}
\end{enumerate}

As can be seen, the criteria can lead to dubious cases, for which additional
checks have to be performed and a discussion of special cases can be found in
Section~\ref{sec:res}. For the curves that do not pass the QC because
$\sigma_{\rm prec} > 10$~mmag or because they failed any other of the QC levels
described previously, we still attempted a constancy assessment, although with
less stringent results:

\begin{enumerate}
\item{if no clear pattern was present in the light curve, and if the various
indicators described in the previous section were compatible with a normal
distribution, we validated the SPSS candidate as {\em likely constant} and kept
it in the SPSS sample; in this case, we set the allowed amplitude of the
variation, if present, as $A < 3~\sigma_{\rm SPSS}  \pm \sigma_{\rm prec}$;}
\item{if some coherent pattern was present, the star was judged {\em suspect
variable}; in that case the amplitude was assigned with the same criteria used for variable
stars and the candidate SPSS was removed from the canditade SPSS list.}
\end{enumerate}


\begin{figure*}
\centering
\includegraphics[width=1.07\textwidth,height=0.95\textheight]{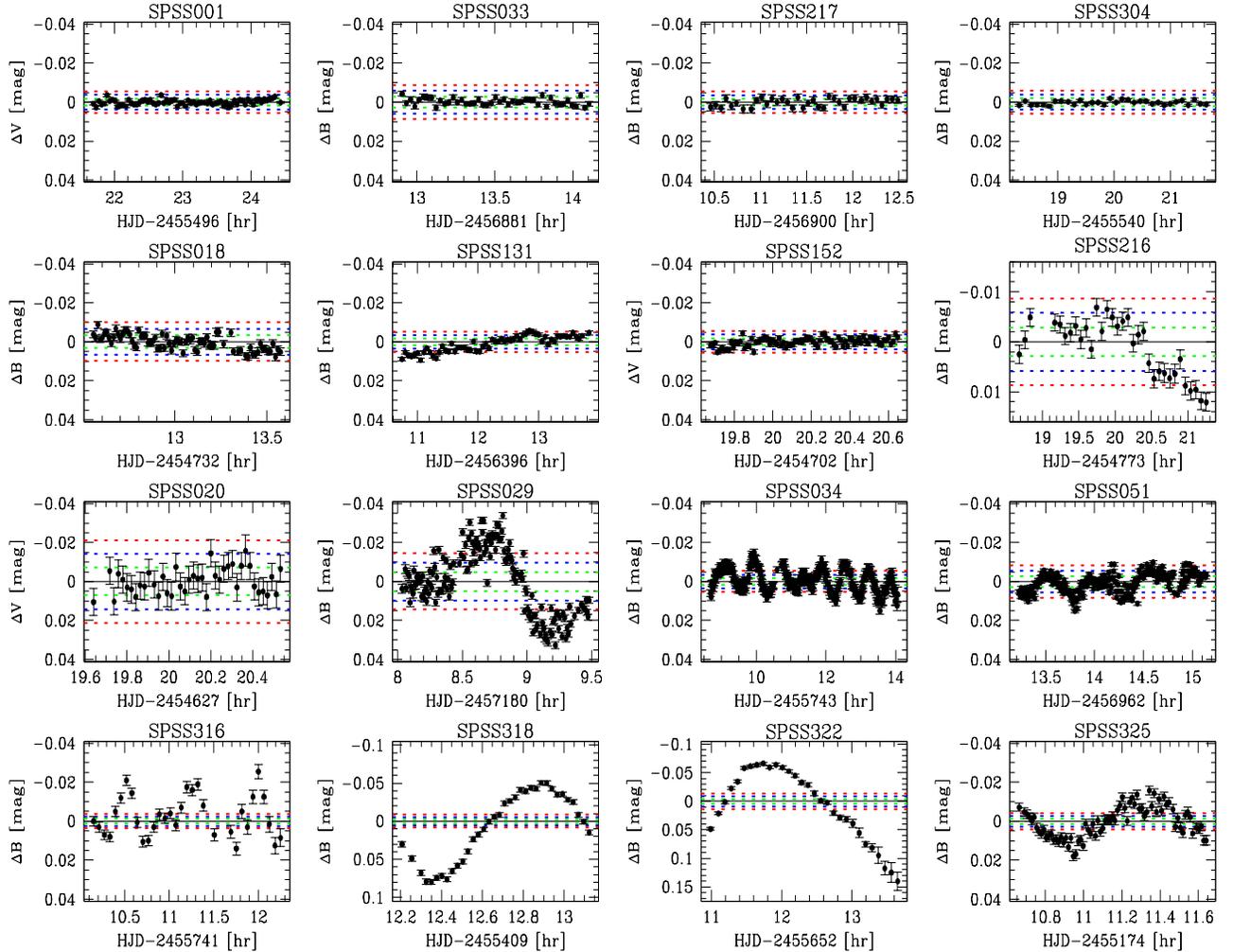} 
\vspace{-8.5cm}
\caption{The best light curve for each monitored SPSS candidate is displayed in
each panel using the same magnitude scale, when possible. The figure is available in its entirety 
in the electronic version of the Journal. Here we show some example of constant stars (the first four panels),
suspect variable stars (panels five to eight, these stars are discussed in Section~\ref{sec:suspectvariables}),
and confirmed variable stars (the remaining panels, these sars are discussed in Section~\ref{sec:variables}).
This is the final zeroed light curve as in the top panel of Figure~\ref{fig:LCcompleteexample}. 
The differential magnitudes are in the sense SPSS minus
supercomparison (see text for more details), i.e., negative values mean that the SPSS becomes brighter. 
The $\pm$1, 2, and 3 $\sigma_{\rm prec}$ thresholds are marked by colored dotted lines in each panel.} 
\label{fig:LCall}
\end{figure*}

\subsection{Additional criteria}
\label{sec:longterm}

We originally planned and started an additional constancy monitoring
campaign on longer timescales (see Paper~I) with the goal of observing
each SPSS candidate 4 times per year, for 3 years. However, because we mostly
obtained the observing time each semester at six different facilities with
normal time applications\footnote{With the exceptions of: NTT for which we
obtained once the ESO large programme status; Calar Alto for which we obtained a
few years of granted time; and TNG for which we obtained a medium-sized
programme allocation once.} it was not practically possible to obtain regularly the
required sampling. We therefore stopped our long-term campaign, also because it
became apparent that Gaia itself will be able to perform this kind of
monitoring excellently on the SPSS sample.

We instead used the absolute night points (see Section~\ref{sec:obsred}) of our
absolute photometry campaign, that are presently being analyzed (a paper is
in preparation). We have at least three independent night points per SPSS
candidate, and in some cases they cover up to 6 years or more for a single SPSS.
Therefore, in some difficult cases we could use absolute night points to
investigate suspicious trends found in our short-term light curves or claims of
long-term variability found in the literature. Some examples of this kind of
additional checks can be found in the discussions in the next section.
Additionally, a thorough literature search for each of the candidate SPSS presented
here was carried out.


\section{Results}
\label{sec:res}

We were able to perform the constancy assessment on a total of
162 SPSS candidates. As a
result, 150 SPSS candidates were validated against short-term variability, 
and only 12 were rejected (8 being clearly identified as variable stars, 
and 4 being classified as suspect variable stars).

We needed to either reject or retain each SPSS, therefore for some uncertain cases we had to
take decisions, motivated by more than the simple criteria described in the
previous section. We discuss in the following the most difficult cases, both to
illustrate the assessment procedure, and to provide useful information for
future studies of these stars. The best light curves obtained for each SPSS are
displayed in Figure~\ref{fig:LCall}, and the outcome of the assessment procedure
is summarized in Table~3, along with some other useful information.

\subsection{Control cases}

The first three SPSS candidates are the {\em Pillars}, three pure Hydrogen WDs 
adopted by \citet{bohlin95} as fun\-da\-men\-tal calibrators (see also Paper~I). We
expect these extremely well studied stars to be constant, and therefore we
observed them as control cases, to verify that with our criteria they
turned out as constant stars. Indeed, even for GD~153 (SPSS~003) that was
observed in a windy night with high and variable seeing and some veils, we found
$\sigma_{\rm prec}$=0.005~mag and no sign of variability. We conclude that even in
the presence of veils our method can be applied.


\subsection{Constant stars}

Of the 126 candidate SPSS that successfully passed all the QC
criteria and were judged constant, five deserve to be discussed. 

A slight trend was observed for GRW+705824 (SPSS~015), especially towards the
end of the light curve (see Figure~\ref{fig:LCall}). Even if the trend was within
$\pm$5~mmag, this could be an indication of longer term variability. We note that
veils appeared during the light curve, and the seeing got worse towards the end
of the series. In spite of all these effects, the curve is of very high quality
($\sigma_{\rm prec}$=0.0017~mag). Only recently, \citet{bohlin15} found some weak
indication of a 0.004~mag/yr, variability after monitoring it from 1986 to 1991.
We thus checked our absolute night points, spanning 3 years (from 2009 to
2012), and we found no significant variation within 0.0065~mag. 

Similarly, the well studied Feige~110 (SPSS~023) shows a weak trend along the
time series, still  contained within $\pm$5~mmag, possibly because the sky was
veiled during the curve acquisition. In the literature, no previous detection of
variability was found. Our absolute night points show large variations in the B
and R bands over one year, but there appear to be problems in the data quality and analysis
for this particular set. On the other hand, the corresponding V magnitude appears constant
within 0.007~mag. Therefore, we accept this SPSS candidate. 

Another similar case is 1812095 (SPSS~037), with a weak trend contained in the
$\pm$5~mmag range, clear sky observations, and no sign of variability in the
literature. Our absolute photometry data contain only one reliable night point
so we cannot use it as an additional constraint.

A different problem was apparent for WD~2028+390 (SPSS~203), where the only
available curve was probably interrupted for the nitrogen refilling of the instrument
and therefore a small jump appears between the two branches of the light curve. Nevertheless, the
SPSS candidate appears constant within 4~mmag and the curve is of good quality.

Finally, SDSS~J125716+220059 (SPSS~355) shows a trend, entirely contained
within  $\pm$10~mmag, that is most probably caused by the high (from $\simeq$5
to 7") and variable seeing, with thin clouds, during observations. No other
curves were obtained for SPSS~355. No sign of variability was found in the
literature, and our absolute photometric measurements are always contained
within a few mmag. So we decided to retain this SPSS candidate.


\subsection{Likely constant stars}

We retained as likely constant 24 SPSS candidates
that formally failed some QC level, or did not completely fulfil our constancy
criteria. 

One example of likely constant stars are SPSS candidates for which only one
of the field comparison stars passed all QC steps, generally because the other comparison stars 
were too faint (S/N ratio below 100). In most cases, including one or more faint
reference stars did not improve significantly the quality of the curve, while in
a few cases, including a second comparison stars with S/N ratio below 100
provided better results. Examples of these SPSS are: EG~21 (SPSS~005), LTT~9491
(SPSS~022), SA~107-544 (SPSS~032), LTT~1020 (SPSS~039), WD~1105-048 (SPSS~121),
WD~1211-169 (SPSS~129), G~16-20 (SPSS~134), WD~0501-289 (SPSS~170),G~179-54 (SPSS~192), 
WD~0123-262 (SPSS~220), and GJ~507.1 (SPSS~335)\footnote{For SPSS~335 all the available comparison stars failed all QC steps.}. 
For all these stars we did not detect any clear sign of variability and no 
variability indication was found in the literature or in our absolute photometry night
points. Of these, SPSS~121 was found to be a magnetic star \citep{aznar04} in a
wide binary system \citep{koester09,zuckerman14}, SPSS~129 is not a white dwarf 
being a G8IV star \citep{kharchenko09}, and SPSS~134 is suspected
to be member of a wide binary system \citep{zapatero04}. These findings do not pose significant danger
to the use of these candidate SPSS, so they were retained in the Gaia SPSS grid.

Another example concerns time series having too few points (less than 30) or
lasting less than one hour, like: Feige~56 (SPSS~026), HD~121968 (SPSS~030),
G~184-20 (SPSS~140), WD~1918+725 (SPSS~142), WD~1319+466 (SPSS~246), WD~1637+335
(SPSS~254), WD~1034+001 (SPSS~308), SDSS~J233024.89--000935.1 (SPSS~321), 
and LP~885-23 (SPSS~352). Each star suffered from various
other drawbacks like low S/N ratio in the SPSS as well, or veils and other
weather problems. In any case, the remaining points generally showed no clear
sign of variability and no variability was found in the literature or in our
absolute photometry. \citet{zapatero04} noted that SPSS~140 is part of a wide
binary system. SPSS~254 was also reported as non-variable by \citet{mcgraw77},
\citet{dolez91}, and \citet{gianninas05}.

One star, WD~2216-657 (SPSS~276), did not pass the criteria because the S/N
ratio of the SPSS was lower than 100 in the only curve available. However, the
curve precision was acceptable ($\sigma_{\rm prec}$=0.008~mag) and no sign of
coherent trends or variability was detected. Our absolute photometry is stable
within 0.015~mag and no indication of variability was found in the literature
either. 

Finally, three stars showed coherent patterns in their light curves, although
they roughly remained within the actual variability limit of 0.01~mag. The first
one, LTT~9239 (SPSS~021), a widely used flux
standard, was observed in good conditions and our absolute
photometry did not show any significant variation. 
The second, WD~0943+441 (SPSS~180), was considered non-variable by
\citet{kepler95} who found it constant within 0.0051~mag. The third WD~0954-710
(SPSS~181), was observed with ROSS2 when there was a problem with the dichroic,
no sign of variability was found in the literature, and our absolute photometry
is all contained within 0.004~mag over 6 years.


\subsection{Suspect variable stars}
\label{sec:suspectvariables}

\begin{figure}
\centering
\includegraphics[width=0.7\columnwidth]{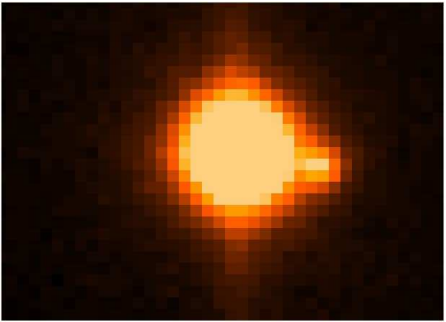} 
\caption{One of the V band images of BD+284211 (SPSS~018) obtained at Calar Alto
with CAFOS showing the presence of a red companion at a distance of about 3".} 
\label{fig:spss018}
\end{figure}

Similarly to the likely constant case, some
candidates SPSS were classified as suspect variable stars either because the data did not
formally pass the QC but some significant trends were revealed, or because we
identified significant trends, but without finding a definite maximum or minimum
in the light curve. Four candidate SPSS were judged suspect
variables and removed from the SPSS candidate list.

The first one is BD+284211 (SPSS~018), a star present in the \citet{landolt07}
catalogue and a widely used CALSPEC standard \citep{bohlin07}. The trend we
observe has an amplitude of almost 0.010~mag in the B band, and although this trend
is still formally within 3~$\sigma_{\rm prec}$, we clearly do not sample the entire period: the variation is therefore 
likely to exceed our criteria. In addition, there are hints in the literature that this
star might be peculiar in many respects. It has a 5~mag fainter red companion at
about 3" \citep{massey90} as shown also by our data in Figure~\ref{fig:spss018},
and it has variable emission in H$_{\alpha}$ with emission lines
\citep{herbig99,latour13}. It is a suspect binary based on its measured infrared
excess \citep{ulla98} and has X-ray emission \citep{lapalombara14,latour15}.

The second one is WD~1327-083 (SPSS~131). This star was classified as non variable by 
\citet{giovannini98}.
Nevertheless, we observe a clear variation of about 0.015 mag in a high 
quality curve lasting more than 3~h and our absolute photometry data show a variation 
of ~0.04 mag in B band in slightly less than one year. The period observed in the light 
curve is too long to be compatible with the typical periods of pulsations for ZZ Ceti stars: 
perhaps SPSS~131 could be a multiple object.  

The spectroscopic binary G~190-15 (SPSS~152) was classified as single-lined spectroscopic binary by
\citet{latham02} and later as double-lined specrtoscopic binary by \citet{halbwachs14}. The level of optical
variability was uncertain in the past. Our best light curve shows a possible
trend that is contained within $\pm$5~mmag and could thus in principle be ignored as done for
a few stars in the previous sections. The curve quality was quite good. However,
our absolute photometry, spanning 3 months, showed a variation of 0.03~mag in B,
that is above our limit of 0.01~mag. 

The last star, WD~0009+501 (SPSS~216), is a magnetic WD, the first one
discovered below 100~kGauss \citep{mccook99}. While we could find no study reporting
light curves or variability detections, we observe a variation of about
0.013~mag in a high-quality light curve, lasting approximately 2.5 hours. The
first part of the light curve is quite flat and then the curve declines, so  we
could not detect any clear maximum or minimum. No apparent technical problems
could explain the trend, but we prefer to avoid using this star in the Gaia SPSS
grid.


\subsection{Confirmed variable stars}
\label{sec:variables}

Eight of the SPSS candidates in our original list (reported in Paper~I) are now known to be
variable and were rejected from the candidate SPSS list. 

The first one is BD+174708 (SPSS~020), a well known and widely used CALSPEC
standard that was also selected as target by the ACCESS rocket mission
\citep{access}. It was shown to brighten by 0.04~mag over a period of 5 years by
\citet{bohlin15}. Our absolute photometry data also show a variation of 0.03
magnitudes in the B, V, and R filters over a period of 7 years. Even if our
short-term data do not show any variation in the monitoring period of 1~hour, we
clearly cannot use it as a Gaia SPSS. 

SA~105-448 (SPSS~029) is another widely used CALSPEC standard and one of the
photometric standard stars in the \citet{landolt83} catalogue. Our best light
curve shows a very definite variability pattern with an amplitude of
0.034$\pm$0.005~mag in B band, compatible with $\delta$~Scuti type pulsations.
We do see in the LC one minimum and one maximum separated by roughly 0.5 hours. 
Among the numerous studies employing this star
as a standard, we could find no indication of variability. Unfortunately we have
only one absolute night point of non-optimal quality. We found data for this star in NSVS
\citep[Northern Sky Variability Survey,][]{wozniak04}. A preliminary analysis of the NSVS data
shows a variation with a period of roughly one hour and amplitude $\simeq$0.03 mag in the V band. 

Another star that was already shown to be variable (see also Paper~I) is
1740346 (SPSS~034). We publish here a longer light curve with respect to that
paper (6~hours), showing various maxima and minima. Its maximum amplitude is
0.015$\pm$0.002~mag in the B band, with various periods and amplitudes. It was
proposed to be a $\delta$~Scuti star by \citet{marinoni11}.

\begin{figure*}
\centering
\includegraphics[width=0.3\textwidth,height=5.5cm]{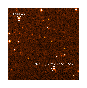} 
\includegraphics[width=0.3\textwidth,height=5.5cm]{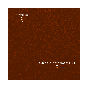} 
\includegraphics[width=0.3\textwidth,height=5.5cm]{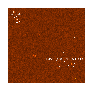} 
\caption{Finding charts of the three comparison stars found
variable during our survey. In all charts, North is up and East is left.
The field of view is $\simeq$ 2', 2', and 7' from left to right respectively.} 
\label{fig:posrefs}
\end{figure*}

Unluckily, another candidate SPSS that was selected to be in common with the targets
of the ACCESS mission also turned out to be variable: HD~37725 (SPSS~051). Its
light curve shows an unmistakable variation over a timescale of roughly half an
hour or less, and a maximum amplitude of 0.015$\pm$0.003~mag in B band. Our
light curve covers almost two hours and a clear global brightening in the
variation pattern is visible. The star is an A3~V and there is evidence of
multiple periods, therefore it is likely a $\delta$~Scuti star. 

Among the several SPSS selected from the SEGUE \citep[The Sloan Extension for Galactic
Understanding and Evolution,][]{beers04} sample to fill in a few
under-represented spectral types, four showed clear signs of variability, with
amplitudes ranging from 0.018$\pm$0.001 to 0.079$\pm$0.003~mag. The first
one, SDSS J164024.18+240214.9 (SPSS~316), was classified as a candidate blue
horizontal branch star by \citet{xue08}, but not confirmed by spectroscopy. 
Due to the evidence of multiple periods in the light curve, it could be a variable of the $\delta$~Scuti type.
For this star, we found some data in NSVS, CSS \citep[Catalina Sky Survey]{catalina}, and WISE \citep[Wide-field Infrared Survey Explorer]{wise}.
However the data have not sufficient quality to draw any firm conclusion.

The second one, SDSS~J224204.16+132028.6 (SPSS~318), was classified as a blue
straggler by \citet{xue08}. Again its light curve is compatible with a $\delta$~Scuti family pulsator with a clear 
minimum and maximum separated by $\simeq$40 minutes. From a preliminary analysis of data available in CSS, we confirm a variation with an amplitude of
roughly 0.05 mag and a period of $\simeq$50 minutes.

The third one, SDSS J122241.66+422443.6 (SPSS~322), has been recently included in the list of
RR~Lyrae discovered in the SDSS-Pan-STARRS-Catalina footprint \citep{abbas14}. 

The last one, SDSS~J000752.22+143024.7 (SPSS~325), has no specific mention of variability in the
literature, but it has low amplitude pulsation and a
period compatible with those of a $\delta$~Scuti star.

Finally, we note that -- even if we did not obtain any useful data -- SDSS J204738.19-063213.1
(SPSS~317) was reported to be an RR Lyrae star by  \cite{abbas14}. Therefore this SPSS was 
excluded from our candidate list.


\begin{figure}
\centering
\includegraphics[width=1.0\columnwidth,height=11cm]{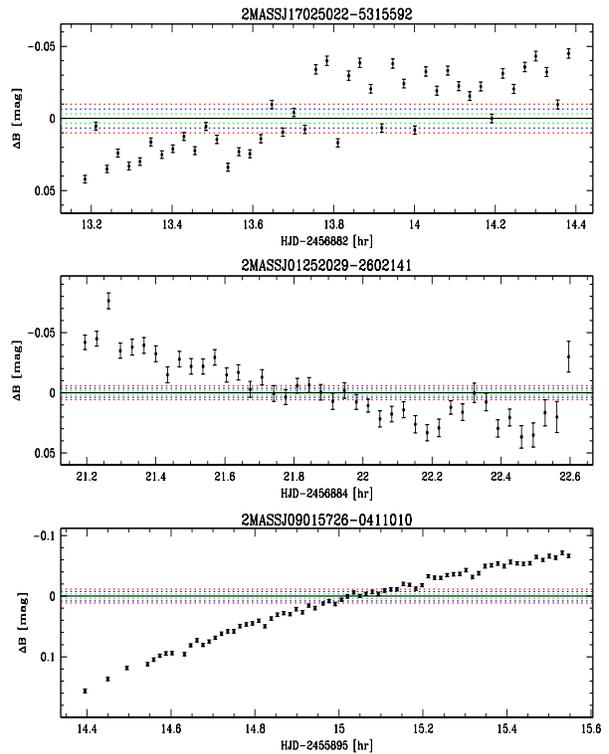} 
\caption{Light curves of the three comparison stars found
variable during our survey.} 
\label{fig:refs}
\end{figure}

\subsection{Variable comparison stars}

Among the field stars that could be used as comparison, we found three that
were variable.

The first one was initially chosen as comparison star for the
light curve of WD~1659--531 (SPSS~196), as shown in the first panel of
Figure~\ref{fig:posrefs}, and later rejected because of its clear variability
pattern. Its light curve is shown in the upper panel of Figure~\ref{fig:refs}.
It was identified as 2MASS~J17025022-5315592, but no
further information was found in the literature. In our light curve, lasting ~1.2 hours,
we could not observe any clear maximum or minimum but the range of the variation
is at least 0.1 mag in B band.

The second one was a comparison star for
WD~0123--262 (SPSS~220), with its position and light curve reported in
Figures~\ref{fig:posrefs} and \ref{fig:refs}, respectively.
The star was identified as 2MASS~J01252029-2602141, but also in this case 
no further information was found in the literature. We observe a variation of about 0.1 mag
in B band, with a possible double minimum towards the end of the curve. The CSS data show a typical RR-Lyrae 
light curve with an amplitude of $\simeq$0.5 mag and a period of $\simeq$0.57 days. 
The double minimum we observed in our data is caused by the typical luminosity bump 
associated with the early shock phase in RRab variables.

The last one was one of the comparison stars for the light curve of
WD~0859--039 (SPSS~307), as shown in Figures~\ref{fig:posrefs} and
\ref{fig:refs}. It was identified as 2MASS~J09015726-0411010, an
eclipsing binary of W~Ursae~Majoris type, with a period of 0.420775~days by the NSVS.
In our curve we observed a variation of about 0.25 mag in B band.


\section{Conclusions}
\label{sec:conclusion}

We presented the results of our extensive short-term (1--2~hrs, see also
Figure~\ref{fig:duration}) constancy monitoring campaign of 162 candidate SPSS
from the list reported in Paper~I, that needed this kind of monitoring. A total of
approximately 143 observing nights -- on a grand-total of more than 515
dedicated to the Gaia SPSS grid ground-based campaigns -- were dedicated to the
short-term constancy monitoring. We present in this paper the best available
curve for each of the monitored SPSS candidates, but on average we obtained
approximately 3 curves for each candidate SPSS. The typical precision
of the light curves was 4~mmag (see also Figure~\ref{fig:sigma}).

We presented and discussed the criteria adopted to assess the constancy or
variability of the candidate SPSS. When the candidate SPSS were judged constant
or likely constant, because no clear variability pattern was present and the
maximum amplitude allowed by our data was estimated as 10~mmag or less, they
were retained in the Gaia SPSS candidate list. Otherwise, the candidate SPSS
were rejected from the list. In all doubtful cases, we made extensive use of the
available literature data and of our absolute photometry measurements (that will
be published in a forthcoming paper) to reach a decision. 

In conclusion, of the 162 monitored SPSS candidates, 150 were validated against short-term variability, 
and only 12 were rejected: eight were found to be variable stars, and four were 
classified as suspected variables. 
Among the confirmed or suspected variable stars, there are some widely used flux
standards from the CALSPEC set, i.e., BD+174708, SA 105-448, 1740346, and HD
37725. 

\section*{Acknowledgments}

We would like to acknowledge the financial support of the INAF (Istituto Nazionale di
Astrofisica) and specifically of the Bologna and Roma Observatories; of ASI (Agenzia
Spaziale Italiana) under contract to INAF: ASI 2014-049-R.0 dedicated to ASDC, 
and under contracts to INAF: ARS/96/77, ARS/98/92, ARS/99/81, I/R/32/00, I/R/117/01, 
COFIS-OF06-01,ASI I/016/07/0, ASI I/037/08/0, ASI I/058/10/0, ASI 2014-025-R.0, 
ASI 2014-025-R.1.2015 dedicated to the Italian participation to the Gaia DPAC (Data Analysis and Processing Consortium).
This work was supported by the MINECO (Spanish Ministry of Economy) - FEDER through grant 
ESP2014-55996-C2-1-R and MDM-2014-0369 of ICCUB (Unidad de Excelencia 'Mar\'\i a de Maeztu').
We acknowledge financial support from the PAPIIT project IN103014. 

The survey presented in this paper relies on data obtained at
ESO (proposals 182.D-0287, 093.D-0197, 094.D-0258), 
Calar Alto (proposals H08-2.2-041, H10-2.2-042, F11-2.2-033), 
TNG (proposals AOT17\_3, AOT19\_14, AOT20\_41, AOT21\_1, AOT29\_Gaia\_013/id7), 
NOT (proposals 49-013, 50-012),
Loiano (15 accepted proposals starting from June 2007), 
San Pedro M\'artir (7 accepted proposals starting from October 2007), 
REM (proposals AOT16\_16012, AOT17\_17012, AOT18\_18002, AOT29\_Gaia\_013/id10), and
TJO (proposal 11010).

We warmly thank the technical staff of the San
Pedro M\'artir, Calar Alto, Loiano, La Silla NTT and REM, Roque de Los
Muchachos TNG and NOT, and Montsec TJO observatories.
Based on observations made with the Italian Telescopio Nazionale Galileo (TNG) operated on the island of La Palma by 
the Fundación Galileo Galilei of the INAF (Istituto Nazionale di Astrofisica) at the Spanish Observatorio del 
Roque de los Muchachos of the Instituto de Astrofisica de Canarias.
Based on observations made with the Nordic Optical Telescope, operated by the Nordic Optical Telescope Scientific 
Association at the Observatorio del Roque de los Muchachos, La Palma, Spain, of the Instituto de Astrofisica de Canarias.
Based on observations collected at the Observatorio Astron\'omico Nacional at San Pedro M\'artir, B.C., Mexico.
We made use of the following softwares and online databases (in alphabetical
order): 2MASS, AAVSO, CALSPEC, Catalina Sky Survey (CSS), CataXcorr, ESO-DSS, IRAF, NSVS, SAOImage DS9, SDSS and SEGUE,
SExtractor, SIMBAD, SuperMongo and WISE. 
The CSS survey is funded by the National Aeronautics and Space
Administration under Grant No. NNG05GF22G issued through the Science
Mission Directorate Near-Earth Objects Observations Program.  The CRTS
survey is supported by the U.S.~National Science Foundation under
grants AST-0909182 and AST-1313422.
This publication makes use of data products from the Wide-field Infrared Survey Explorer, which is a joint project of the University of California, 
Los Angeles, and the Jet Propulsion Laboratory/California Institute of Technology, funded by the National Aeronautics and Space Administration.

\begin{landscape}
\begin{table}
\label{tab:results}
\caption{Results. The columns contain: 
{\em (1)} the SPSS internal ID; 
{\em (2)} the name of the star; 
{\em (3)} and {\em (4)} the star coordinates compiled in Paper~I; 
{\em (5)} the spectral type compiled in Paper~I;
{\em (6)} the instrument used; 
{\em (7)} photometric band of the best observed time series;
{\em (8)} the sky quality during the acquisition of our best time series;
{\em (9)} indicates whether the following column is an upper ($\leq$ ) or lower ($\geq$)
limit or a determination (=) of the amplitude;   
{\em (10)} the minimum tested amplitude in case of constant stars, or the maximum observed amplitude in case of variable stars;  
{\em (11)} the precision of the light curve used (see text);
{\em (12)} the number of comparison stars used for the SPSS light curve production;
{\em (13)} specifies whether the star will be used as a Gaia SPSS; 
{\em (14)} any additional comment.
The table is available in its entirety in the electronic version of the Journal and at CDS. Here we show a portion to illustrate its contents.}
\begin{tabular}{@{}lllllclclllcll@{}}
\hline
SPSS   &  Name &   RA (J2000)    & Dec (J2000) & Sp       & Instrument & Band  & Sky     & lim   &  A     & $\sigma_{\rm prec}$  & \# Comp.   &  Verdict  & Notes  \\
       &       &  (hh:mm:ss)     & (dd:mm:ss)  & Type     &            &       & Quality &       & (mag)  & (mag)                &  Stars     &           & \\
\hline
001 & G~191-B2B       & 05:05:30.61    & +52:49:51.95	 & DA0   & LaRuca  & V & good     & $\leq$ &  0.0014  &  0.0019  &   3    & Accepted  &  Pillar, control case \\
002 & GD~71           & 05:52:27.63    & +15:53:13.37	 & DA1   & LaRuca  & B & good     & $\leq$ &  0.0026  &  0.0034  &   3    & Accepted  &  Pillar, control case \\
003 & GD~153          & 12:57:02.33    & +22:01:52.52	 & DA1   & BFOSC   & B & windy    & $\leq$ &  0.0049  &  0.0050  &   2    & Accepted  &  Pillar, control case. Bad seeing \\
005 & EG~21           & 03:10:31.02    & --68:36:03.39   & DA3   & EFOSC2  & B & veiled   & $\leq$ &  0.0056  &  0.0011  &   1    & Accepted  &  Likely constant. Only one comparison  \\
008 & LTT~3218        & 08:41:32.56    & --32:56:34.90   & DA	 & EFOSC2  & B & good     & $\leq$ &  0.0032  &  0.0038  &   2    & Accepted  &  ---  \\
009 & LTT~2415        & 05:56:24.74    & --27:51:32.35   & G	 & ALFOSC  & B & good     & $\leq$ &  0.0042  &  0.0036  &   2    & Accepted  &  ---  \\
010 & GD~108          & 10:00:47.37    & -00: 33:30.50   & B	 & LaRuca  & B & veiled   & $\leq$ &  0.0065  &  0.0082  &   3    & Accepted  &  ---  \\
011 & Feige~34        & 10:39:36.74    & +43:06:09.25	 & DO	 & BFOSC   & B & veiled   & $\leq$ &  0.0032  &  0.0037  &   3    & Accepted  &  ---  \\ 
012 & LTT~4364        & 11:45:42.92    & --64:50:29.46   & DQ6   & ROSS2   & g & good     & $\leq$ &  0.0032  &  0.0037  &   4    & Accepted  &  ---  \\ 
013 & Feige~66        & 12:37:23.52    & +25:03:59.87	 & O	 & LaRuca  & B & veiled   & $\leq$ &  0.0063  &  0.0079  &   2    & Accepted  &  ---  \\ 
014 & HZ~44           & 13:23:35.26    & +36:07:59.51	 & O	 & DOLORES & B & cloudy   & $\leq$ &  0.0031  &  0.0055  &   2    & Accepted  &  ---  \\ 
015 & GRW+705824      & 13:38:50.47    & +70:17:07.62	 & DA3   & BFOSC   & B & veiled   & $\leq$ &  0.0018  &  0.0017  &   2    & Accepted  &  Possible trend, but A$<$5mmag  \\ 
017 & LTT~7987        & 20:10:56.85    & --30:13:06.64   & DA4   & EFOSC2  & B & good     & $\leq$ &  0.0028  &  0.0027  &   2    & Accepted  &  ---  \\ 
018 & BD+284211       & 21:51:11.02    &  +28:51:50:36   & Op	 & BFOSC   & B & good     & $\geq$ &  0.0088  &  0.0033  &   3    & Rejected  &  Suspect variable \\ 
020 & BD+174708       & 22:11:31.37    &  +18:05:34.17   & F8	 & ROSS    & V & varying  & $\leq$ &  0.0065  &  0.0071  &   2    & Rejected  &  Long term variability \\
021 & LTT~9239        & 22:52:41.03    & --20:35:32.89   & F	 & ALFOSC  & B & good     & $\leq$ &  0.0152  &  0.0030  &   2    & Accepted  &  Likely constant, possible trend\\
022 & LTT~9491        & 23:19:35.44    & --17:05:28.40   & DB3   & EFOSC2  & B & veiled   & $\leq$ &  0.0159  &  0.0039  &   1    & Accepted  &  Likely constant. Only one comparison  \\
023 & Feige~110       & 23:19:58.40    & --05:09:56.21   & O	 & BFOSC   & B & veiled	  & $\leq$ &  0.0026  &  0.0029  &   5    & Accepted  &  Possible trend, but A$<$5mmag \\ 
024 & HILT~600        & 06:45:13.37    &  +02:08:14.70   & B1	 & LaRuca  & B & good     & $\leq$ &  0.0034  &  0.0040  &   3    & Accepted  &  ---  \\ 
025 & AGK+81266       & 09:21:19.18    &  +81:43:27.64   & O	 & BFOSC   & B & good     & $\leq$ &  0.0017  &  0.0026  &   5    & Accepted  &  ---  \\ 
026 & Feige~56        & 12:06:47.23    &  +11:40:12.64   & B5p   & BFOSC   & B & veiled   & $\leq$ &  0.0332  &  0.0108  &   2    & Accepted  &  Likely constant. Less than 1 hour  \\ 
027 & Feige~67        & 12:41:51.79    &  +17:31:19.75   & O	 & BFOSC   & B & veiled   & $\leq$ &  0.0031  &  0.0030  &   5    & Accepted  &  ---  \\ 
028 & SA~105-663      & 13:37:30.34    & --00:13:17.37   & F	 & BFOSC   & B & veiled   & $\leq$ &  0.0068  &  0.0079  &   2    & Accepted  &  ---  \\  
029 & SA~105--448     & 13:37:47.07    & --00:37:33.02   & A3	 & BFOSC   & B & good     & =	   &  0.0338  &  0.0048  &   2    & Rejected  &  Variable \\ 
\hline
\end{tabular}
\end{table}
\end{landscape}


\bsp

\label{lastpage}

\end{document}